# Quantum Search on Structured Problems

Lov K. Grover, 3C-404A Bell Labs, 600 Mountain Avenue, Murray Hill NJ 07974 *(lkgrover@bell-labs.com)*

**Summary**

This paper shows how a basic property of unitary transformations can be used for meaningful computations. This approach immediately leads to search-type applications, where it improves the number of steps by a square-root - a simple minded search that takes $N$ steps, can be improved to approximately $\sqrt{N}$ steps. The quantum search algorithm is one of several immediate consequences of this framework. Several novel search-related applications are presented.

**1. Introduction** Several interesting problems in computer science can be looked upon as search problems. There are two categories of such problems. First, where the search depends on data in memory - this is the database search kind of problems. Alternatively, the search could be based on a function known in advance - many NP-complete problems and cryptography problems can be expressed in this form. For example the SAT problem of NP-completeness asks whether there exists a combination of binary variables that satisfies a specified set of Boolean equations - this can be looked upon as a search of the state space of the binary variables. In cryptography, the well-known 56-bit DES code (Data Encryption Standard) can be cracked by an exhaustive search of $2^{56}$ items [BBHT][Phone].

It aroused considerable interest when it was shown that it was possible to improve upon the obvious classical bound for exhaustive search by resorting to quantum mechanics [Search][BBHT] - the intuitive reason for this improvement was that quantum mechanical systems can be in multiple states and simultaneously explore different regions of configuration space. This improved the number of steps by a square-root, i.e. a simple minded search that takes $N$ steps, could be improved to approximately $\sqrt{N}$ steps. The quantum search algorithm was derived using the Walsh Hadamard (W-H) transform and it appeared to be a consequence of the special properties of this transform. Subsequently [Gensrch] showed that similar results are obtained by substituting *any* unitary transformation in place of the W-H transform. This means that a variety of unitary transformations could be used in place of the W-H transform and this leads to algorithms for several different problems. This paper describes the approach of [Gensrch] and shows how it can be extended to solve various structured problems.

**2. Framework** A function $f(x)$, $x = 0, 1, \ldots (N-1)$, is given which is known to be zero for all $x$ except the single point $x = t$, the goal is to find $t$ ($t$ for target). The obvious classical technique of searching by looking at the $N$ values of $x$, one by one, would clearly take $O(N)$ steps.

Assume that we have at our disposal a unitary transformation $U$ that acts on a system with $N$ basis states. First map each value of $x$ to a basis state and start with the system in the basis state $s$ ($s$ for start). If we apply $U$ to $s$, the amplitude of reaching $t$ is $U_{ts}$, and if we were to make a measurement that projects the system into a unique basis state, the probability of getting the right basis state would be $|U_{ts}|^2$. It would, therefore, take $\Omega\left(\frac{1}{|U_{ts}|^2}\right)$ repetitions of this experiment before a single success. This section shows how it is possible to reach state $t$ in only $O\left(\frac{1}{|U_{ts}|}\right)$ steps. This leads to a sizable improvement in the number of steps if $|U_{ts}| \ll 1$.

Denote the unitary operation that inverts the amplitude in a single state $x$ by $I_x$. In matrix notation this is the diagonal matrix with all diagonal terms equal to 1, except the $xx$ term which equals $-1$.

$v_x$ denotes the column vector which has all terms zero, except for the $x^{th}$ term which is unity.

Consider the following unitary operator: $Q \equiv -I_s U^{-1} I_t U$, since $U$ is unitary $U^{-1}$ is equal to the *adjoint,* i.e. the complex conjugate of the transpose of $U$. We first show that $Q$ preserves the two dimensional vector space spanned by the two vectors: $v_s$ and $(U^{-1} v_t)$ (note that in the situation of interest, when $U_{ts}$ is small, these two vectors are almost orthogonal).

First consider $Qv_s$. By the definition of $Q$, this is: $-I_s U^{-1} I_t U v_s$. Note that $v_x v_x^T$ is an $N \times N$ square matrix all of whose terms are zero, except the $xx$ term which is 1. Therefore $I_t \equiv I - 2v_t v_t^T$ & $I_s \equiv I - 2v_s v_s^T$, it follows:

(1) $\quad Qv_s = -(I - 2v_s v_s^T) U^{-1} (I - 2v_t v_t^T) U v_s = -(I - 2v_s v_s^T) U^{-1} U v_s + 2(I - 2v_s v_s^T) U^{-1} (v_t v_t^T) U v_s$

Using the facts: $U^{-1} U = I$ and $v_s^T v_s \equiv 1$, it follows that:

(2) $\quad Qv_s = v_s + 2(I - 2v_s v_s^T) U^{-1} (v_t v_t^T) U v_s$.

Simplify the second term of (2) by the following identities: $v_t^T U v_s \equiv U_{ts}$ & since $U$ is unitary, its inverse is equal to its *adjoint (*the complex conjugate of the transpose) $v_s^T U^{-1} v_t \equiv U_{ts}^*$.

(3) $\quad Qv_s = v_s(1 - 4|U_{ts}|^2) + 2U_{ts}(U^{-1} v_t)$

Next consider the action of the operator $Q$ on the vector $U^{-1} v_t$. Using the definition of $Q$ (i.e. $Q \equiv -I_s U^{-1} I_t U$) and carrying out the algebra as in the computation of $Qv_s$ above, this yields:

(4) $\quad Q(U^{-1} v_t) \equiv -I_s U^{-1} I_t U(U^{-1} v_t) = -I_s U^{-1} I_t v_t = I_s U^{-1} v_t$.

Writing $I_s$ as $I - 2v_s v_s^T$ and as in (3), $v_s^T U^{-1} v_t \equiv U_{ts}^*$:

(5) $\quad Q(U^{-1}v_t) = (U^{-1}v_t) - 2v_s v_s^T (U^{-1}v_t) = (U^{-1}v_t) - 2U_{ts}^* v_s .$

It follows that $Q$ transforms any superposition of $v_s$ & $(U^{-1}v_t)$ into another superposition of the two vectors, thus preserving the two dimensional vector space spanned by $v_s$ & $(U^{-1}v_t)$. (3) & (5) may be written as:

(6) $\quad Q \begin{bmatrix} v_s \\ U^{-1}v_t \end{bmatrix} = \begin{bmatrix} (1 - 4|U_{ts}|^2) & 2U_{ts} \\ -2U_{ts}^* & 1 \end{bmatrix} \begin{bmatrix} v_s \\ U^{-1}v_t \end{bmatrix}$

It follows as in [BBHT], that if we start with $v_s$, then after $\eta$ repetitions of $Q$ we get the superposition $a_s v_s + a_t (U^{-1} v_t)$ where $a_s \cong \cos(2\eta |U_{ts}|)$ & $|a_t| \cong |\sin(2\eta |U_{ts}|)|$ if $|U_{ts}| \ll 1$. If $\eta = \dfrac{\pi}{4|U_{ts}|}$, then we get the superposition $U^{-1} v_t$; from this with a single application of $U$ we can get $v_t$. Therefore in $O\left(\dfrac{1}{|U_{ts}|}\right)$ steps, we can start with the $s$-state and reach the target state $t$ with certainty.

**3. Quantum Operations** The interesting feature of the analysis of section 2 is that $U$ can be *any* unitary transformation, whatsoever. Clearly, it can be used to design algorithms where $U$ is a transformation on the qubits in a quantum computer - the object of this paper is to present some such applications. Quantum mechanical operations that can be carried out in a controlled way are unitary operations that act on a small number of qubits in each step. It is possible to design a variety of quantum mechanical algorithms using just a few elementary quantum mechanical operations. Two of the elementary unitary operations needed are: the W-H transformation operation and the selective inversion of the amplitudes of certain states.

A basic single bit operation in quantum computing is the operation $M$ - this is represented by the following matrix: $M \equiv \dfrac{1}{\sqrt{2}} \begin{bmatrix} 1 & 1 \\ 1 & -1 \end{bmatrix}$, i.e. a bit in the state 0 is transformed into a superposition: $\left(\dfrac{1}{\sqrt{2}}, \dfrac{1}{\sqrt{2}}\right)$. Similarly a bit in state 1 is transformed into $\left(\dfrac{1}{\sqrt{2}}, -\dfrac{1}{\sqrt{2}}\right)$. In a system in which the states are described by $n$ bits (it has $N = 2^n$ possible states) we can perform the operation $M$ on each bit independently in sequence thus changing the state of the system. The state transition matrix representing this operation will be of dimension $2^n \times 2^n$. Consider a case when the starting state is one of the $2^n$ basis states, i.e. a state described by an $n$-bit binary string with some 0s and some 1s. The result of performing the operation $M$ on each bit will be a superposition of states described by all possible $n$-bit binary strings with the amplitude of each state being $\pm 2^{-\frac{n}{2}}$. To deduce the sign, observe that from the definition of the matrix

$M$, i.e. $M \equiv \frac{1}{\sqrt{2}}\begin{bmatrix} 1 & 1 \\ 1 & -1 \end{bmatrix}$, that the phase of the resulting configuration is changed only when a bit that was previously a 1 remains a 1 after the transformation is performed. Hence if $\bar{x}$ be the $n$-bit binary string describing the starting state and $\bar{y}$ the $n$-bit binary string describing the resulting string, the sign of the amplitude of $\bar{y}$ is determined by the parity of the bitwise dot product of $\bar{x}$ and $\bar{y}$, i.e. $(-1)^{\bar{x} \cdot \bar{y}}$. This transformation is the W-H transformation [DJ]. This operation (or a closely related operation called the Fourier Transformation [Factor]) is one of the things that makes quantum mechanical algorithms more powerful than classical algorithms and forms the basis for most significant quantum mechanical algorithms.

The other transformation we will need is the selective inversion of the phase of the amplitude in certain states. Unlike the W-H transformation and other state transition matrices, the probability in each state stays the same since the square of the absolute value of the amplitude in each state stays the same. The following is a realization based on [BBHT]. Assume that there is a binary function $f(x)$ that is either 0 or 1. Given a superposition over states $x$, it is possible to design a quantum circuit that will selectively invert the amplitudes in all states where $f(x) = 1$. This is achieved by appending an ancilla bit, $b$ and considering the quantum circuit that transforms a state $|x, b\rangle$ into $|x, f(x) XOR\, b\rangle$ (such a circuit exists since, as proved in [Revers], it is possible to design a quantum mechanical circuit to evaluate any function $f(x)$ that can be evaluated classically). If the bit $b$ is initially placed in a superposition $\frac{1}{\sqrt{2}}(|0\rangle - |1\rangle)$, this circuit will invert the amplitudes precisely in the states for which $f(x) = 1$, while leaving amplitudes in other states unchanged.

**4. Summary of Applications** As mentioned in section 1, the search problem is the following: a function $f(x), x = 0, 1, \ldots (N-1)$, is given which is known to be non-zero at certain values of $x$; the task is to find one such value. No structure is known for $f(x)$ except for what is explicitly mentioned in the specific problems $(4.1)\ldots(4.7)$. $N$ is assumed to be a power of 2, say $N = 2^n$. There is a one-to-one correspondence between the $N$ values of $x$ and the respective states of an $n$-bit register. States corresponding to values of $x$ for which $f(x)$ is non-zero, are referred to as $t$-states.

(4.1) Assume that $f(x) = 0$ everywhere except for a single value of $x$. This is the standard problem of exhaustive search.

(4.2) As in (4.1), there is a single point, $t$, where $f(x)$ is non-zero. Some information about $t$ is available in the following form - another $n$ bit word, $r$, is given which is known to differ from $t$ in at most $k$ out of the $n$ bits.

(4.3) There are multiple points ($t$-states) at which $f(x)$ is non-zero, it is required to find any one of these. Some structure to the problem is specified in the following form. We are given a certain unitary transformation $U$ &

multiple *s*-states so that $U_{ts}$ for any *s* & any *t* are the same. The setting of this subsection is abstract, (4.4), (4.5) & (4.7) apply the framework of this subsection to concrete problems.

(4.4) As in (4.3), there are multiple points at which $f(x)$ is non-zero, it is required to find any one of these. However, unlike (4.3), no further structure to the problem is given.

(4.5) There is a single point *t* where $f(x)$ is non-zero. Some information about *t* is available in the form of *l n*-bit strings, each of which differ from *t* in exactly *k* out of *n* bits.

(4.6) $f(x) = 0$ everywhere except at the unique point $x = t$, it is required to find *t*. Also, as in (4.3), we are given a unitary transformation *U* and multiple *s*-states. However, unlike (4.3), $U_{ts}$ for various *s*-states & various *t*-states are not all identical.

The analysis of section 2 extends the power of quantum search so that it can be used with an arbitrary unitary transform *U*, but only with a single *s* and single *t*-state. (4.3) extends it to multiple states, but in a restricted way. This derivation extends to multiple *s*-states. It is still not known how to handle multiple *t*-states that are not exactly symmetric.

(4.7) This problem illustrates how the abstract techniques discussed earlier can be applied to solve an actual problem. This problem was first discussed by Eddie Farhi & Sam Gutmann [Structure].

Two functions $f(x, y)$ & $g(x)$ are defined on the domain $x = 0, 1, ...(N-1)$, $y = 0, 1, ...(N-1)$. $f(x, y)$ is zero everywhere except at the unique point $(t_1, t_2)$, $g(x)$ is non-zero at *M* values of *x* including $x = t_1$ (here $M \ll N$). The problem is to find $t_1$ & $t_2$. Classically this problem would take $\Omega(NM)$ steps. The algorithm of (4.1), without using the function $g(x)$, would take $O(N)$ steps. The following analysis makes use of the general technique of (4.3) to develop an $O(\sqrt{NM})$ step algorithm. Several variants of this problem are also considered.

**4.0 The Approach** The following general approach is made use of in each of the next 7 sub-sections - (4.1)…(4.7).

There are $N = 2^n$ states, represented by *n* qubits, the task is to get the system into some target state(s) *t* at which $f(x)$ is non-zero. A unitary transform *U* and the initial state *s* are selected and $U_{ts}$ is calculated. It then follows by section 2 that by $\frac{\pi}{4|U_{ts}|}$ repetitions of the operation sequence $-I_s U^{-1} I_t U$, followed by a single application of *U*, the initial state *s* is transformed into the final state *t*.

**4.1 Exhaustive Search** Assume that $f(x) = 0$ everywhere except at a single point *t*. The object is to find *t*.

As mentioned in the first paragraph of the introduction, there are several important problems in computer science for which there no solution is known, except exhaustive search.

**Solution** For the W-H transform, described in section 3, $U_{ts}$ between *any* pair of states $s$ & $t$ is $\pm\frac{1}{\sqrt{N}}$. Therefore we can start with any state $s$ and the procedure of (4.0) gives us an algorithm requiring $O\left(\frac{1}{|U_{ts}|}\right)$ steps, i.e. $O(\sqrt{N})$ steps.

In case $s$ be chosen to be the $\bar{0}$ state, then the operation sequence $Q = -I_{\bar{0}}WI_tW$ leads to the standard quantum search algorithm based on the *inversion about average* interpretation [Gensrch] (note that $W^{-1} = W$). To see this write $I_{\bar{0}}$ as $I - 2v_{\bar{0}}v_{\bar{0}}^T$. Therefore for any vector $\bar{x}$: $-WI_{\bar{0}}W\bar{x} = -W\left(I - 2v_{\bar{0}}v_{\bar{0}}^T\right)W\bar{x} = -\bar{x} + 2Wv_{\bar{0}}v_{\bar{0}}^TW\bar{x}$. It is easily seen that $Wv_{\bar{0}}v_{\bar{0}}^TW\bar{x}$ is another vector each of whose components is the same and equal to $A$ where $A \equiv \frac{1}{N}\sum_{i=0}^{N-1} x_i$ (the average value of all components). Therefore the $i^{th}$ component of $-WI_{\bar{0}}W\bar{x}$ is simply: $(-x_i + 2A)$. This may be written as $A + (A - x_i)$, i.e. each component is as much above (below) the average as it was initially below (above) the average, which is precisely the *inversion about average*.

In case $s$ be chosen to be a state different from $\bar{0}$, the dynamics is still very similar to the standard quantum search algorithm; however, the *inversion about average* interpretation no longer applies.

**4.2 Search when an item *near* the desired state is known:** This problem is similar to (4.1), i.e. $f(x) = 0$ except at the single point $t$. The difference from (4.1) is that some information about the solution, $t$, is available in the following form: another $n$ bit word, $r$, is specified - $t$ is known to differ from $r$ in at most $k$ of the $n$ bits.

Such a problem would occur in any situation when we had some prior information about the solution, this information could come either from prior knowledge or from a noisy data-transmission.

***Solution*:** The effect of the constraint is to reduce the size of the solution space. One way of making use of this constraint, would be to map this to another problem and then exhaustively search the reduced space using (4.1). However, such a mapping would involve additional overhead. This section presents a different approach which carries over to more complicated situations as in (4.5).

Instead of choosing $U$ as the W-H transform, as in (4.1), in this section $U$ is tailored to the problem under consideration. The starting state $s$ is chosen to be the specified word $r$. The operation $U$ consists of the following unitary transformation $\begin{bmatrix} \sqrt{1-\frac{k}{n}} & \sqrt{\frac{k}{n}} \\ \sqrt{\frac{k}{n}} & -\sqrt{1-\frac{k}{n}} \end{bmatrix}$, applied to each of the $n$ qubits. Calculating $U_{ts}$, it follows that

$$|U_{ts}| = \left(1 - \frac{k}{n}\right)^{\frac{n-k}{2}} \left(\frac{k}{n}\right)^{\frac{k}{2}} \text{ and } \log|U_{ts}| = \frac{n}{2}\log\frac{n-k}{n} - \frac{k}{2}\log\frac{n-k}{k}.$$ The technique described in (4.0) can now be used - as in (4.1), this consists of repeating the sequence of operations $-I_s U I_t U$, $O\left(\frac{1}{|U_{ts}|}\right)$ times, followed by a single application of the operation $U$ (note that, as in (4.1), $U^{-1} = U$).

The size of the space being searched in this problem is approximately $\binom{n}{k}$ which is equal to $\frac{n!}{n-k!k!}$. Using Stirling's approximation: $\log n! \approx n\log n - n$, it follows that $\log\binom{n}{k} \approx n\log\frac{n}{n-k} - k\log\frac{k}{n-k}$, comparing this to the number of steps required by the algorithm, we find that the number of steps in this algorithm, as in (4.1), varies as the square-root of the size of the solution space being searched.

**4.3 Multiple $s$ & $t$ states with the same $U_{ts}$:** $f(x)$ is non-zero at $\beta$ values of $x$, i.e. there are $\beta$ $t$-states. Some structure of the problem is specified in the following form. Assume that we have at our disposal a unitary transform $U$ and $\alpha$ $s$-states such that $U_{ts}$ between *any* $t$-state and *any* $s$-state is the same. The object is to find one of the $t$-states. This is accomplished by transforming the system into a superposition so that there is an equal amplitude in each of the $t$-states and zero amplitude elsewhere. After this, a measurement is made that projects the system into one of its basis states, this gives a $t$-state.

The problem considered in this subsection is abstract in the sense $U$ is an arbitrary unitary transformation. (4.4), (4.5) and (4.7) apply this to concrete problems.

**Solution:** The approach is similar to the exhaustive search problem of (4.1). However, the analysis of section 2 has to be redone with the following three changes:

(a) The starting state instead of being $v_s$, is the superposition $\frac{1}{\sqrt{\alpha}} \sum_{a=0}^{\alpha-1} v_{s_a}$ - the amplitude in all $s_a$ states is equal to $\frac{1}{\sqrt{\alpha}}$, and zero everywhere else. Assuming $\alpha$ to be a power of 2 ($\alpha \equiv 2^a$), such a superposition can be easily created by the following procedure. Start with an $a$ bit system with all bits in the 0 state. Do a W-H transform on the $a$ bit system and then carry out a mapping from the $2^a$ states to the $s$-states.

(b) The operations $I_s$ & $I_t$ invert the amplitudes in *all* $s$-states & *all* $t$-states, respectively.

(c) It can then be shown by an analysis similar to section 2, that after $O\left(\frac{1}{\sqrt{\alpha\beta}|U_{ts}|}\right)$ operations of $-I_s U^{-1} I_t U$ followed by a single application of $U$, the system reaches a superposition so that the amplitude is equal in all

$\beta$ $t$-states and is zero everywhere else. Note that the number of operations is smaller by a factor $\sqrt{\alpha\beta}$ as compared to the situation when there were single $s$-states & single $t$-states (as in (4.1)).

**4.4 Problem** $f(x)$ is non-zero at $\beta$ values of $x$, equivalently there are $\beta$ $t$-states - the task is to find one of these.

This is the problem of exhaustive search when there are multiple ($\beta$) solutions. A classical search would take an average of $O\left(\frac{N}{\beta}\right)$ steps to find a solution. This section presents an $O\left(\sqrt{\frac{N}{\beta}}\right)$ step quantum mechanical algorithm.

**Solution** By the definition of the W-H transform in section 3, $W_{t\bar{0}}$ for any $t$ is the same. Therefore if we choose $s$ as the $\bar{0}$ state, then it follows by (4.3) that after $O\left(\sqrt{\frac{N}{\beta}}\right)$ repetitions of $-I_s W I_t W$ followed by a single application of $W$, the system reaches a superposition such that the amplitude is equal in all the $t$ states and zero everywhere else. Note the following three points regarding this scheme:

- As in (4.3), the operation inverts the phase for all $\beta$ $t$-states.
- The above implementation requires $\beta$ to be known in advance.
- The search time is $\sqrt{\beta}$ faster than the exhaustive search algorithm of (4.1).
- It is necessary to choose $s$ as the $\bar{0}$ state, this is different from (4.1) where $s$ could be arbitrary.

**4.5 Problem** $f(x) = 0$ except at the single point $t$. Some information about $t$ is available in the form of $\alpha$ $n$-bit strings, each of which differs from $t$ in *exactly k* bits.

This is in some sense the dual of (4.4). In that case there were multiple $t$-states but a single $s$-state, while in this problem there are multiple $s$-states and a single $t$-state. This kind of problem could occur in extracting a signal out of multiple noisy transmissions.

**Solution** Let the $\alpha$ specified states be the $s$-states. Initialize the system to a superposition of these states by the process described in (4.3)(a). After this, apply the unitary transform $U$ which applies the following unitary operation

$\begin{bmatrix} \sqrt{1-\frac{k}{n}} & \sqrt{\frac{k}{n}} \\ \sqrt{\frac{k}{n}} & -\sqrt{1-\frac{k}{n}} \end{bmatrix}$ to each qubit. As in (4.2), $|U_{ts}| = \left(1-\frac{k}{n}\right)^{\frac{n-k}{2}} \left(\frac{k}{n}\right)^{\frac{k}{2}}$ and $\log|U_{ts}| = \frac{n}{2}\log\frac{n-k}{n} - \frac{k}{2}\log\frac{n-k}{k}$ for all $s$-states. Also, since each of the $s$-states differ from $t$ in exactly the same number of bits implies that $U_{ts}$ has the same sign for all $s$-states. The framework of (4.3) can now be used - this yields an algorithm that is $\sqrt{\alpha}$ times faster than that of (4.2).

**4.6 Multiple *s*-states & a single *t*-state such that $U_{ts}$ between various *s*-states & *t* is not identical** $f(x) = 0$ everywhere, except at the single point *t*. As in (4.3), some structure to the problem is specified via $U$. Assume that we have at our disposal a unitary transform $U$ and various *s*-states are specified. However, unlike (4.3), $U_{ts}$ between various *s* and various *t*-states are *not* exactly equal. This case is qualitatively different from all of those considered so far because the analysis of section 2 or the modified analysis of (4.3), does not directly apply.

This extends the noisy data-transmission problem of (4.5), to the case where there are multiple states specified that are close to the solution state but differ from it in varying number of bits ((4.5) required each of the given states to differ from the solution in *exactly* the same number of bits).

**Solution** Assume the number of *s*-states to be $\alpha$, further assume that $\alpha$ is a power of 2, i.e. $\alpha \equiv 2^a$. Consider $V$ to be a unitary matrix which is a product of 3 simpler unitary matrices, i.e. $V \equiv V_1 V_2 U$ and $V^{-1} \equiv U^{-1} V_2^{-1} V_1^{-1}$.

$V_1$ is a W-H transformation on *a* bits and $\alpha$ states where $\alpha = 2^a$,

$V_2$ maps the $\alpha = 2^a$ states generated by $V_1$ onto the respective *s*-states in *N*-dimensional state space,

$U$ is the available unitary transform on the *N* states.

Let the initial state be the $\bar{0}$ state. As a result of $V_1 V_2$, the amplitude in each of the $s_a$ states: $s_0, s_1 \ldots s_{\alpha-1}$, becomes $\frac{1}{\sqrt{\alpha}}$; after $U$, the amplitude in the *t*-state is $\frac{1}{\sqrt{\alpha}} \sum_{a=0}^{\alpha-1} U_{ts_a}$. By (4.0), it follows that after $\frac{\sqrt{\alpha}}{\left|\sum_{a=0}^{\alpha-1} U_{ts_a}\right|}$ repetitions of $Q = -I_{\bar{0}} V^{-1} I_t V$ followed by a single application of $V$, the amplitude in the target state becomes O(1).

Many of the problems in the previous subsections can be seen to be particular cases of this. For example in case $U_{ts_a}$ are all equal, say to $u$, then the number of iterations becomes: $\frac{1}{|u|\sqrt{\alpha}}$. The algorithm and bound of (4.5) immediately follow from this.

**4.7 Two dimensional search** Two functions $f(x, y)$ & $g(x)$ are defined on the domain $x = 0, 1, \ldots (N-1)$, $y = 0, 1, \ldots (N-1)$. $f(x, y)$ is zero everywhere except at the unique point $(t_1, t_2)$, $g(x)$ is non-zero at $M$ values of *x* including $x = t_1$ (here $M \ll N$). The problem is to find $t_1$ & $t_2$.

Classically this problem would take $\Omega(NM)$ steps. The algorithm of (4.1), without using the function $g(x)$ would take $O(N)$ steps. This section presents an $O(\sqrt{NM})$ step algorithm. For a different analysis, along with a proof that the algorithm of this section is within a constant factor of the fastest possible algorithm, see [Structure].

Several variations of this problem are also briefly considered, these demonstrate how the techniques discussed in this paper can be applied to real problems.

**Solution** First consider the function $g(x)$, $x = 0, 1, \ldots(N-1)$. By executing the algorithm of (4.4), with the *t*-states as the non-zero values of $g(x)$, it is possible for the system to reach a superposition such that at each point at which $g(x)$ is non-zero, the amplitude is $\frac{1}{\sqrt{M}}$ and the amplitude is zero everywhere else. This is accomplished by a sequence of $O\left(\sqrt{\frac{N}{M}}\right)$ unitary transformations (4.4) - denote this composite unitary operation by $U_1$. Next keep the value of *x* the same and carry out the algorithm of (4.1) on the *N* values of *y* with the *t*-state corresponding to the non-zero value of the function $f(x, y)$. This consists of a sequence of $O(\sqrt{N})$ elementary unitary transformations, denote this composite unitary operation by $U_2$. It follows from (4.1), that as a result of this operation sequence, in case the particle is at $x = t_1$, it is also at $y = t_2$. The amplitude of the system being in the desired state is therefore $\frac{1}{\sqrt{M}}$. By means of the unitary transformation $U = U_1 U_2$, the system starting from a certain initial state reaches the desired state with an amplitude of $\frac{1}{\sqrt{M}}$.

Since $U$ is a sequence of elementary unitary operations, it follows that $U^{-1}$ is a sequence of the adjoints of the same operations in the opposite order and can hence be synthesized. By applying the procedure of section 2, it follows that by repeating the sequence of operations $Q \equiv -I_s U^{-1} I_t U$, $O(\sqrt{M})$ times, the system reaches the target state with certainty.

The total number of steps is given by the number of repetitions of $Q$ (i.e. $O(\sqrt{M})$) times the number of steps required for each repetition, (i.e. $\left(O\left(\sqrt{\frac{N}{M}}\right) + O(\sqrt{N})\right)$) which gives $O(\sqrt{M}) \times \left(O\left(\sqrt{\frac{N}{M}}\right) + O(\sqrt{N})\right) = O(\sqrt{NM})$.

The above analysis easily extends to more general cases. For example consider the case, where the search-space is rectangular instead of square, i.e. the number of possible values for *x* is $N_1$ and for *y* is $N_2$, The number of steps, instead of being $O(\sqrt{NM})$, now becomes $O(\sqrt{N_1}) + O(\sqrt{N_2 M})$. Alternatively consider the case where there is an $\eta$ dimensional space with the same number of points (*N*) in each dimension. Instead of just $f(x, y)$ & $g(x)$, there are now $\eta$ functions: $f(x_1, x_2, \ldots x_\eta)$, $g_1(x_1, x_2, \ldots x_{\eta-1})$, $g_2(x_1, x_2, \ldots x_{\eta-2})$, ..., $g_{\eta-1}(x_1)$ with analogous definitions. It follows by a similar approach that the number of steps is now $O(\sqrt{NM_1 M_2 \ldots M_{\eta-1}})$.

Another variation is when the function $f(x, y)$ is non-zero at multiple points, say $\beta$ points, and it is required to find one of these. In case each of the β points has a different value of $x$, the framework of (4.3) applies. Considering the unitary transformation $U$ as $U_1 U_2$, leads to an algorithm that requires $O\left(\sqrt{\frac{NM}{\beta}}\right)$ steps.

It is not clear how to derive an algorithm when some of the β points have the same value of $x$.

**5. Conclusion & Further Work** [Search] demonstrated how to make use of quantum mechanical properties to develop exhaustive search kinds of algorithms, i.e. algorithms for problems that lacked any structure. [Search] used subtle properties of a particular quantum operation called the W-H transform. Subsequently [Gensrch] extended this so that other quantum operations could be used instead of the W-H transform.

Most interesting problems in computer science are concerned with the structure of problems and how to develop algorithms to take advantage of this structure. [Hogg] has previously suggested heuristic quantum mechanical algorithms for structured problems. This paper has given several examples of structured problems and how search-type algorithms can be extended to solve these ((4.2) through (4.7)) - quantitative closed form bounds were derived for the running time of these algorithms. The extensions have shown how to deal with the situation where there are multiple *s*-states (initial states) and a single *t*-state (target state) ((4.5) and (4.6)). Also, it is possible to deal with multiple *t*-states provided they are exactly symmetric ((4.3) and (4.4)). The next step would be to obtain a general algorithm with multiple *t*-states.